\documentclass[aps,twocolumn,pra,superscriptaddress,showpacs,tightenlines]{revtex4-1}
\usepackage{epsfig,graphicx,times}
\usepackage{amstext}
\usepackage{amsmath}
\usepackage{amssymb}
\usepackage{graphicx}
\usepackage{latexsym}
\usepackage{bm}
\usepackage[colorlinks,citecolor=blue,urlcolor=blue, linkcolor=blue,hyperindex,CJKbookmarks,dvipdfm]{hyperref}

\begin{document}

\title{Escape of  \texorpdfstring{$\alpha$}--particle in inertial confinement fusion}
\author{Kai Li}
\email{likai@csrc.ac.cn}
\affiliation{Institute of Applied Physics and Computational Mathematics,
Beijing 100088, China}
\author{Ke Lan}
\email{ke\_lan@hotmail.com}
\affiliation{Institute of Applied Physics and Computational Mathematics,
Beijing 100088, China}
\date{\today }
\begin{abstract}
Escape of  $\alpha$-particles from a burning or an ignited burning deuterium-tritium (DT) fuel with temperature up to  more than tens  keV  is very important in  inertial  confinement fusion, which can significantly influence not only the hot spot dynamics and the energy gain but also the shielding design in fusion devices. In this paper, we study the $\alpha $-particle escape   from a burning or an ignited burning  DT  fuel by considering the modifications including the $\alpha $-particle stopping by both DT ions and  electrons  with their   Maxwellian average stopping weights, the relativity effect on electron distribution, and the modified Coulomb logarithm of the DT-$\alpha $ particle collisions. As a result of our studies, the escape-effect from our modified model is obviously stronger than those from the traditional models.
A fitted expression is presented to calculate the escape factor in a DT fuel, which can be applied to a burning fuel with temperatures of 1 to 150 keV and areal densities of 0.04 to 3 g/cm$^2$ with an accuracy within $\pm0.02$. Finally, we discuss the $\alpha $-particle escape-effect on the hot-spot dynamics and the thermonuclear energy gain by comparing the results with  escape factors from different models.
\end{abstract}

\maketitle


\section{Introduction}

In Inertial Confinement Fusion (ICF) \cite{ICF1,ICF2,ICF3}, a spherical encapsulated equimolar deuterium-tritium (DT) fuel is imploded isotropically  either directly by lasers ~\cite{Campbell2017} or indirectly by X-ray radiation converted inside a hohlraum from lasers \cite{Haan2011,Lan2014,Lan2016,indirect3} to a very high velocity of about 3 $\times 10^7$ to 4 $\times 10^7$ cm/s,  so that the fuel is highly compressed by rocket effect~\cite{rocket3,rocket2,rocket} to extreme density, temperature and pressure  under the spherical convergent effect, and a  hot spot with an areal density of about 0.3 g/cm$^2$ and a temperature of about 5 keV is formed in the center, triggering the nuclear fusion  of the main fuel and resulting in a significant thermonuclear energy gain. Thermonuclear ignition and burn of a plasma occurs when self-heating by fusion products $\alpha$-particles exceeds all energy losses such that no further external heating
is necessary to keep the plasma in the burning state. That is to say, the $\alpha $-particle self-heating is a prerequisite of ignition and burn of the fuel plasma.
However, not all fusion produced $\alpha $-particles can be deposited in the fuel, because part of them can escape. Calculation of  the escape of the $\alpha$-particles can significantly influence not only the theoretical studies on the ignition condition and the fusion energy  gain but also the designs of the ignition targets. Especially, for an ignited burning fuel, its temperature can be up to more than tens  keV, while the escape-effect is larger at a higher temperature and it can seriously influences the  shielding material and configuration design of future fusion reactors.

Up to now, various theoretical models have been developed for the  $\alpha$-particle escape, such as
the Krokhin and  Rozanov model~\cite{Krokhin1973}  (KR model), the Fraley model~\cite{Fraley1974}, the  Atzeni and Meyer-ter-Vehn~\cite{Atzeni2004} model (AM model), and the latest Zylstra and Hurricane model
  (ZH model) ~\cite{Zylstra2019}.
 The first model to estimate the escape-effect of $\alpha$-particles was proposed by Krokhin and  Rozanov~\cite{Krokhin1973} and then improved by Atzeni and Meyer-ter-Vehn~\cite{Atzeni2004}, which are widely used till now ~\cite{Ribeyre2011,Park2014,Hurricane2014,Cheng2015,Hurricane2016,Huang2017,Cheng2018}.
In KR model, it ignored the deceleration of $\alpha $%
-particles induced by DT ions  and  assumed a fuel with uniform  density and
temperature. As a result,  the ratio of the escaped energy to the initial
energy of the $\alpha $-particles, called as the escape factor of   $\alpha $-particle  in this paper, is obtained analytically as %
\begin{equation}
\eta_A =\left\{
\begin{array}{c}
\frac{1}{4\tau }-\frac{1}{160\tau ^{3}}\text{, \ \ }\tau \geq \frac{1}{2} \\
\\
1-\frac{3\tau }{2}+\frac{4}{5}\tau ^{2}\text{, }\tau <\frac{1}{2}%
\end{array}%
\right. \text{ ,}  \label{Ratio_Ref}
\end{equation}%
for the  spherical DT fuel. Here, $\tau =\frac{\rho R}{\rho l_{\alpha}}$ is  stopping depth of  $\alpha$-particle in the fuel, $\rho$ is density, $\rho R$ is   areal density of the fuel, $l_{\alpha}$ is  $\alpha$-particle range, and $\rho l_{\alpha}$ is  areal density range of the $\alpha$-particles. In AM model, it gives  a fitted formula for $\rho l_{\alpha}$ as
\begin{equation}
\rho l_{\alpha}=\frac{0.025T^{\frac{5}{4}}}{1+0.0082T^{\frac{5}{4}}}
\label{rho_l_Atzeni}
\end{equation}
by taking the DT ions into account, and  uses Eq.~(\ref{Ratio_Ref}) to calculate $\eta_A$. Here, $\rho l_{\alpha}$ is in units of g/cm$^2$ and $T$, the temperature of the fuel, is in units of keV.
According to a comparison among the KR model, the Fraley model and the AM model
 by Ref.~\cite{Zylstra2019}, the AM model has the longest calculated range of the  $\alpha$-particles and thus the lowest heating efficacy, which results from the calculation at densities more relevant to modern ICF designs in the AM model.
In the latest ZH model, it used the modern stopping-power theories    and gave an expression for calculating   the  $\alpha$-particle escape from a fuel of temperature limited in the range of 1$\sim$10 keV.

Notice that it uses   Eq.~(\ref{Ratio_Ref})  to calculate the $\alpha$-particle escape factor in AM model, which involves the $\alpha$-particle  stopping   induced only by electrons.
Thus, it needs a consistent expression for the $\alpha$-particle escape factor, which    considers the  stopping contributions not only from electrons but also from  the DT ions.
In addition, the Coulomb logarithms for the DT-$\alpha$ collisions in   AM model is obtained by using the  thermal velocity of the DT ions as  the relative velocity between these two kinds of particles, which needs to be modified when an $\alpha$-particle velocity  is much higher than the thermal velocity  of the DT ions. Furthermore, for an ignited burning fuel with temperatures up to tens to hundred  keV, the relativity effect on the electron velocity distribution   can not be neglected in calculating the   $\alpha$-particle  escape.

In this paper, based on the classical two-body Coulomb collision model, we will give  modifications on the $\alpha $-particle escape factor, including the $\alpha $-particle stopping by both DT ions and  electrons  with their   Maxwellian average stopping weights,
the electron stopping weight modified by the relativity effect,
and the relative velocity effect   on  the Coulomb logarithm of DT-$\alpha $ particle collisions. In the modifications,   we also consider the quantum effect on the Coulomb logarithm of  DT-$\alpha $ particle collisions  though it is small in the ICF range.
  As we know, the $\alpha $-particle escape factor  is strongly connected with the time and space dependent plasma status of the fuel. However, to compare the escape factors from our model with those from the published models, we will consider  a one-dimensional spherical DT fuel with uniform temperature and density and  give a fitted expression of the $\alpha$-particle escape factor. Because the $\alpha $-particle escape can seriously influence the self-heating and the energy gain of a DT fuel, so we will further  discuss and compare the $\alpha $-particle escape-effect on the hot-spot dynamics by using    escape factors from  different models.

This paper is organized as follows. In Sec. II, based on the classical two-body Coulomb collision model, we study the $\alpha $-particle escape from a DT fuel with temperatures up to more than tens keV by considering the Maxwellian averaged stopping weights of both DT ions and  electrons,
the electron stopping weight modified by the relativity effect, and  the relative velocity effect and the  quantum effect on  Coulomb logarithm of the DT-$\alpha $ particle collisions.
In Sec. III,  we   compare  the $\alpha$-particle escape factor  from our model with those from  previous published models for  a DT fuel. In Sec. IV, we study the escape-effect of the $\alpha$ -particles on  hot-spot dynamics of a   DT fuel   by using different escape factors. Finally, we will present a summary in Sec. V.

\section{Stopping of \texorpdfstring{$\alpha$}--particle}

In a burning fuel or an ignited burning  fuel with strong self-heating, the temperature can range from a few keV to more than hundred keV. Inside such a fuel, the stopping of the 3.54 MeV $\alpha $-particle is mainly contributed by thermal electrons and  thermal fuel ions via collisions, which  is usually
described by the Coulomb collision model. In this work, we   take the two-body classical elastic Coulomb collisions for a  fully ionized fuel  and truncate the impact parameter at Debye shielding length~\cite{Debye}.
The total energy loss within a path length $ds$ of an $\alpha $-particle with an instantaneous energy $E_{\alpha }$ is calculated as%
\begin{equation}
\frac{dE_{\alpha }}{ds}=-\sum\limits_{j }\left(\frac{dE_{\alpha }}{ds}\right)_j%
\text{ ,}  \label{dE_alpha}
\end{equation}%
where  the subscript $j$ represents the particles of different  species which collide with the $\alpha $-particles, and it refers to either thermal electrons or thermal DT ions for a fully ionized pure DT plasma in the  fuel. We have:
\begin{equation}
\left(\frac{dE_{\alpha }}{ds}\right)_j=\frac{n_{j }(q_{\alpha }q_{j
})^{2}\ln \Lambda _{j }}{8\pi \mu _{j }\varepsilon _{0}^{2}v_{\alpha
}}\int_{0}^{\infty }f_{j }(v_{j })h_j(v_{j })dv_{j }
\text{ ,} \label{dE_j}
\end{equation}
where it is assumed that the species $j $ has an isotropic velocity distribution function of $f_{j }(v_{j })$.
In Eq.~(\ref{dE_j}),  $n $ is the  number density, $q $ is the charge, $v$ is the velocity  and the subscript refers to either  a species  $j$  or an $\alpha$-particle, $\varepsilon_0$ is the permittivity of vacuum, $\ln \Lambda _{j }$ is the Coulomb logarithm of $j $-$\alpha$ collision, and $\mu _{j }=\frac{m_{\alpha
}m_{j }}{m_{\alpha }+m_{j }}$ is the reduced mass of a species $j $ and an $\alpha$-particle, respectively.
Notice that the factor $h(v_{j })$ weights the deceleration or acceleration of
an $\alpha$-particle with $v_{\alpha }$ when it collides with a $j$-particle with $v_{j}$, which is written as %
\begin{equation}
h_j(v_{j })=\left\{
\begin{array}{c}
-\frac{2\mu _{j }}{m_{\alpha }v_{j }}\text{, \ \ \ \ \ \ \ }%
v_{j } > v_{\alpha } \\
\frac{2}{v_{\alpha }}\left( 1-\frac{\mu _{j }}{m_{\alpha }}\right) \text{%
, \ }v_{j }\leq v_{\alpha }%
\end{array}%
\right. \text{.}  \label{H-function}
\end{equation}%
Thus, an $\alpha$-particle is decelerated by a slower $j$-particle with $v_{j } < v_{\alpha }$, while accelerated by a faster $j$-particle with $v_{j } > v_{\alpha }$. However, the deceleration weight   can be different from the acceleration weight. From  Eq.~(\ref{H-function}),
the deceleration weights are remarkably higher than   the  acceleration weights for either DT ions or electrons. Especially, the deceleration weight of a slow  electron  can be at least four orders higher than the acceleration weight of a fast electron, as presented in Fig. ~\ref{distribution}.
From  Eq.~(\ref{dE_j}), the $\alpha $-particle stopping is strongly connected with the
$\alpha$-particle velocity, the   velocity distribution of species  $j$ and the   Coulomb logarithm of $j $-$\alpha$ collision.
In the following, we consider following modifications on the $\alpha $-particle stopping inside an ignited burning fuel with fully ionized DT plasmas: the $\alpha $-particle stopping by both DT ions and  electrons  with their   Maxwellian average stopping weights, the electron stopping weight modified by the relativity effect, and the modified Coulomb logarithm of the DT-$\alpha $ collisions with  the relative velocity effect and the quantum effect, respectively.

\subsection{Maxwell-averaged stopping weights}

Assuming the  $j$ species is in thermal equilibrium with the Maxwellian velocity distribution,  we can obtain the  same result as in Ref.~\cite {Krokhin1973} from Eq. (\ref{dE_j}):
\begin{equation}
\left(\frac{dE_{\alpha }}{ds}\right)_j=\frac{q_{\alpha }^{2}q_{j
}^{2}n_{j }}{4\pi m_{j }\varepsilon _{0}^{2}v_{\alpha }^{2}}\ln
\Lambda _{j }g_{j }(\frac{v_{\alpha }}{v_{\mathrm{th},j }})\text{
.}  \label{dE-Maxwellian}
\end{equation}%
Here, $v_{\mathrm{th},j }=\sqrt{\frac{2T_{j }}{m_{j }}}$ is the thermal velocity of $j $  species at its temperature $T_{j }$, and  $g_{j }$ is a function of the ratio of $v_{\alpha }$ to $v_{\mathrm{th},j }$ with the following expression:
\begin{equation}
g_{j }(x)=\mathrm{erf}(x)-\left( 1+\frac{m_{j }%
}{m_{\alpha }}\right)x\frac{d}{dx}\mathrm{erf}(x)\text{ ,}  \label{g}
\end{equation}%
where $\mathrm{erf}(x)=\frac{2}{\sqrt{\pi }}\int_{0}^{x}\exp (-y^{2})dy$ is
the error function. We call $g_{j }$ as the stopping weight of  $j$ species on the
$\alpha$-particle energy change per unit length, which is  important in calculating  the escape  of $\alpha$-particles.

We consider the $\alpha$-particle stopping  inside a burning fuel or an ignited burning fuel of fully ionized DT plasmas. Transferring inside the DT fuel, an $\alpha$-particle with initial velocity $v_{\alpha}^0$ collides with both  DT ions  and  electrons,   and its velocity varies and can be obviously lower than $v_{\alpha}^0$.
We use $g_{DT}$ and $g_e$ to express the stopping weight of DT ions and electrons, respectively.
For  $v_{\alpha }\ll v_{\mathrm{th},e}$, which is always valid
in the total deceleration process of $\alpha $-particle at $T_{e}\gg 1$ keV,
$g_{e}$ can be written approximately as
\begin{equation}
g_{e}(\frac{v_{\alpha }}{v_{\mathrm{th},e}})\approx\frac{4}{3\sqrt{\pi }}(%
\frac{v_{\alpha }}{v_{\mathrm{th},e}})^{3} \text{ ,}  \label{g-e}
\end{equation}
which is widely used in previous works~\cite%
{Krokhin1973,Fraley1974}.
In the following text, we use this approximation for $g_{e}$.

\begin{figure}[htb]
\centering
\includegraphics[width=0.42\textwidth]{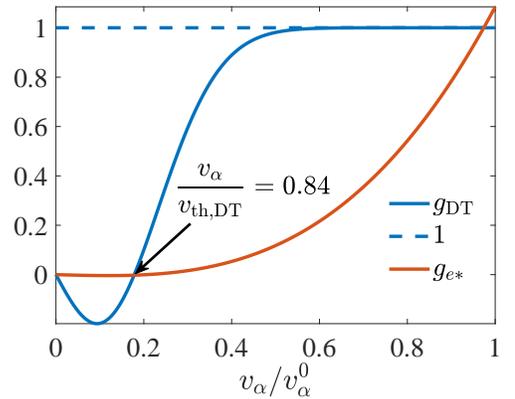}
\caption{(Color online) Variations of $g_{\mathrm{DT}}$ (full blue),  $g_{e*}$ (full red) as $v_{\alpha}/v_{\alpha}^0$ at $T_{\mathrm{DT}}=T_e=100$ keV. The approximation of taking $g_{\mathrm{DT}} = 1$ (dashed blue) is also given for comparison.}
\label{approximation}
\end{figure}

If we neglect  the slow variations of $\ln\Lambda _{\mathrm{DT} }$ and $\ln\Lambda _{e}$,  then the  ratio of the stopping contribution  from DT ions to that from electrons is mainly decided by
$m_eg_{\mathrm{DT}}/( m_{\mathrm{DT}}g_e)$
from Eq.~(\protect\ref{dE-Maxwellian}).
To compare with $g_{\mathrm{DT}}$, we define  $g_{e*}=$ $\frac{m_{\mathrm{DT}}}{m_e}g_e$. Presented in Fig.~\ref{approximation} is the  variations of $g_{\mathrm{DT}}$ and $ g_{e*} $   as $v_{\alpha}/v_{\alpha}^0$ at $T_{\mathrm{DT}}=T_e=100$ keV.  As shown, the electron stopping weight is larger than the DT ion stopping weight for a newly produced $\alpha $-particle with a velocity around $v_{\alpha}^0$. However,  the  stopping weight of DT ions is obviously larger than that of the electrons during almost whole deceleration phase of an $\alpha $-particle, such as with its velocity decreasing from 0.95$v_{\alpha}^0$ to 0.2$v_{\alpha}^0$ at 100 keV. With a velocity of about 0.2$v_{\alpha}^0$, the same as $v_{\mathrm{th, DT}}$ at 100 keV, the $\alpha $-particle  stops deceleration at this temperature. In the following, we discuss the stopping  of $\alpha $-particle by the DT ions and the electrons, respectively.

From  Fig.~\ref{approximation},  $g_{\mathrm{DT}}$ at 100 keV approaches its maximum of 1 at $v_{\alpha} > 0.5v_{\alpha}^0$, but it drops abruptly as decrease of $v_{\alpha}$ at  $v_{\alpha}\lesssim 0.5v_{\alpha}^0$ and  even drops to zero at  $v_{\alpha} = 0.2 v_{\alpha}^0$. In particular, $g_{\mathrm{DT}}$ is negative at $v_{\alpha} < 0.2 v_{\alpha}^0$ from Fig.~\ref{approximation} or at $v_{\alpha }<0.84 v_{\mathrm{th,DT}}$ from Eq.~(\protect\ref{g}), which means that an $\alpha $-particle  with a velocity lower than $v_{\mathrm{th,DT}}$ can even gain energy from DT ions instead of losing energy to DT ions in this case.
Thus,  the stopping condition of $\alpha $-particle  should be defined as $v_{\alpha } = v_{\mathrm{th,DT}}$, instead of $v_{\alpha }=0$. Here, it is worth to mention that an approximation of $g_{\mathrm{DT}}\approx 1$ is often used in previous works~\cite%
{Krokhin1973,Fraley1974,Atzeni2004} with a consideration of $v_{\mathrm{th,DT}}\ll v_{\alpha
}$, which seriously overestimates the $\alpha $-particle stopping power by using  a much bigger  stopping weight of 1 for those $\alpha $-particles with velocities obviously lower than $v_{\alpha}^0$.

Again from Fig.~\ref{approximation},  $g_{e*}$ at 100 keV is obviously larger than $g_{\mathrm{DT}}$  at $  v_{\alpha } > 0.95 v_{\alpha}^0$, while much smaller than $g_{\mathrm{DT}}$  at $0.2 v_{\alpha}^0 \le v_{\alpha } \le 0.95 v_{\alpha}^0$. It means that the stopping   is  mainly dominated by the electrons for the newborn  $\alpha $-particles or when the $\alpha $-particles are in the initial status of  deceleration, while  dominated by the DT ions after the  $\alpha $-particle is significantly decelerated. Thus,  calculation of  $g_e$ is especially important for a newborn  $\alpha $-particle. In addition,  $g_{e*}$ is about zero at $v_{\alpha } < 0.2 v_{\alpha}^0$ at 100 keV from Fig.~\ref{approximation}, which means the electrons have no influence at all for  those seriously decelerated $\alpha $-particles.

\subsection{Electron stopping weight modified by relativity effect}
\begin{figure}[htbp]
\centering
\includegraphics[width=0.45\textwidth, clip]{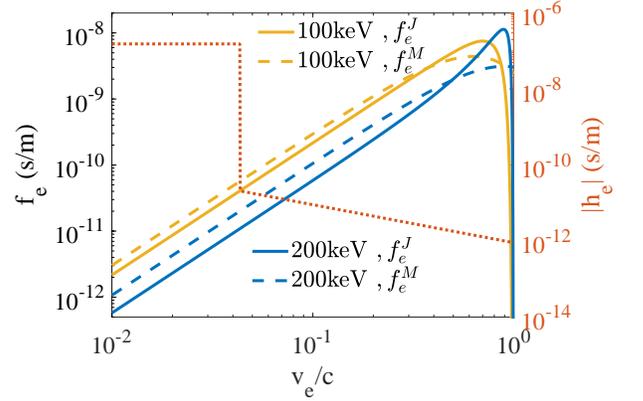}
\caption{(Color online) Absolute value of $h_e(v_e)$ from Eq.(5) (dotted red, right y-axis) and
 comparisons (left y-axis) between the Maxwell distributions~(dashed lines) and
 the Maxwell-J\"{u}ttner distributions~(solid lines) for electrons  at  100 keV (yellow) and 200 keV (blue).  }
\label{distribution}
\end{figure}

As shown in Fig.~\ref{distribution},   the stopping of the $\alpha $-particles  by electrons is dominated by the slow electrons which velocities satisfy $v_{e}\leq v_{\alpha }\ll c$, where $c$ is the velocity of light in vacuum. It means that the dynamics of Coulomb collision between $\alpha $-particles and electrons exempt from the relativity effect, but nevertheless, the number of
electrons with velocities $v_{e}\leq v_{\alpha }$ is very different when the relativity effect is taken into account at a temperature higher than tens keV, which is easily achieved in an ignited burning fuel. Thus, the relativity can influence the stopping via the electron distribution.
Here, we consider the relativity effect and give a modification on the electron stopping weight $g_e$.

In deducing Eq. (\ref{dE-Maxwellian}) from Eq. (\ref{dE_j}), we use the following Maxwell distribution function of electrons without considering the relativity effect:%
\begin{equation}
f_{e}^{M}(v_{e})=\frac{4}{\sqrt{\pi }}v_{\mathrm{th},e}^{-3}v_{e}^{2}\exp
\left( -\frac{v_{e}^{2}}{v_{\mathrm{th},e}^{2}}\right) \text{ .}  \label{f_M}
\end{equation}%
 However, the electron velocity follows the   Maxwell-J\"{u}ttner distribution function when  the relativity effect is taken into consideration~\cite{Juettner1911}:%
\begin{equation}
f_{e}^{J}(v_{e})=\frac{v_{e}^{2}}{\gamma c^{3}K_{2}(\frac{1}{\gamma })}%
\lambda ^{5}\exp \left( -\frac{\lambda }{\gamma }\right) \text{ .} \label{f_J}
\end{equation}%
Here $\gamma =\frac{T_{e}}{m_{e}c^{2}}$, $\lambda
=(1-\frac{v_{e}^{2}}{c^{2}})^{-\frac{1}{2}}$, and $K_{2}$ is the second
order modified Bessel function ~\cite{Abramowitz1988}.
As shown in  Fig.~\ref{distribution},
it has a significant difference between the Maxwell distribution and the Maxwell-J\"{u}ttner distribution  for the slow electrons at 100 keV and 200 keV. At a higher temperature, the difference between the two kinds of electron velocity distributions is bigger.
We can define a relativity factor $\xi$ as:%
\begin{eqnarray}
\xi &=&\frac{\int_0^c f_e^J(v_e)h_e(v_e)dv_e}{\int_0^c f_e^M(v_e)h_e(v_e)dv_e},  \label{relativity_O}
\end{eqnarray}%
and modify  the electron stopping weight $g_e$   as $\xi g_e$ under the relativity effect.
Considering that the  stopping weight of electrons is mainly contributed by the slow electrons
with velocity of $v_e\leq v_\alpha\ll \{v_{\mathrm{th},e},c\}$, we can take an expansion
of the  two distribution functions Eqs. (\ref{f_M}, \ref{f_J}) at $v_{e}=0$, respectively,
as%
\begin{eqnarray}
f_{e}^{M}(v_{e}) &\approx &\frac{4}{\sqrt{\pi }}v_{\mathrm{th}%
,e}^{-3}v_{e}^{2}\text{ ,} \\
f_{e}^{J}(v_{e}) &\approx &\frac{v_{e}^{2}}{\gamma c^{3}K_{2}(\frac{1}{%
\gamma })}\exp (-\frac{1}{\gamma })\text{ .}
\end{eqnarray}%
Further considering that function $h_e(v_e)$ has a constant value in the interval $v_{e}\leq v_{\alpha }$ from Eq. (\ref{H-function}), we have
\begin{equation}
\xi \approx \frac{\sqrt{\pi }v_{\mathrm{th},e}^{3}\exp (-\frac{1}{\gamma })}{%
4\gamma c^{3}K_{2}(\frac{1}{\gamma })} \text{ .}  \label{relativity_general}
\end{equation}
In addition, we also take an expansion of
$K_{2}(\frac{1}{\gamma })$ at $\gamma =0$ by considering $\gamma \ll 1$ at $T_{e}\ll
m_{e}c^{2}\approx 511$ keV. Then the relativity factor can be simplified as
\begin{eqnarray}
\xi &\approx &\frac{\sqrt{2}v_{\mathrm{th},e}^{3}}{4c^{3}(\gamma ^{\frac{3}{2}}+%
\frac{15}{8}\gamma ^{\frac{5}{2}})}  \notag \\
&=&\frac{8}{8+15\gamma }\text{ .}  \label{relativity}
\end{eqnarray}%
Notice that Eq. (\ref{relativity})
depends only on $T_e$. At a very low  $T_e$ with  $\gamma \approx 0$, $\xi$ tends to
1, and the Maxwell-J\"{u}ttner distribution approaches the Maxwell distribution. As $T_e$ increases, the number of electrons with $v_e<v_\alpha$ decreases under both  distributions,
but this effect is more significant under the Maxwell-J\"{u}ttner distribution, as shown in Fig.~\ref{distribution}. It means that
the electron stopping weight   decreases more seriously as $T_e$ increases under the Maxwell-J\"{u}ttner distribution. In other words, the relativity effect of the electrons increases the escape probability of the $\alpha$-particles out of fuel.
For example, $\xi \approx 0.73$ at $\gamma \approx 0.2$ or $T_e$ = 100 keV, which means that relativity effect leads to a decrease of $27\%$ on the electron stopping weight  as compared to the result under the Maxwell distribution.

\subsection{Modifications on Coulomb logarithms}

From Eq. (\ref{dE-Maxwellian}), the $\alpha $-particle stopping is  connected with $\ln \Lambda _{\mathrm{DT}}$ and $\ln \Lambda _{e}$, which are the Coulomb logarithms for collisions between the
$\alpha $-particles and the DT ions and collisions between the $\alpha $-particles and the electrons, respectively.
Generally,  the Coulomb logarithm is calculated with
\begin{equation}
\ln \Lambda _{j}=\ln\frac{\lambda_D}{\max(l_{j,\mathrm{C}}, l_{j,\mathrm{Q}})} \text{ ,}
\label{log_j}
\end{equation}
where
\begin{equation}
\lambda_D=\sqrt{\frac{\varepsilon_0T}{2e^2n}}
\label{l_D}
\end{equation}
is the Debye shielding length,   $l_{j,\mathrm{C}}$ is a velocity-averaged classical impact parameter, and     $l_{j,\mathrm{Q}}$ is a velocity-averaged quantum impact parameter. Here, we assume $T_e=T_{\mathrm{DT}}\equiv T$, $n_e=n_{\mathrm{DT}}\equiv n$, and the factor 2 in Eq. (\ref{l_D}) accounts for the Debye shielding induced by both electrons and DT ions. The  classical impact parameter for a 90$^\circ$ scattering of $j$-$\alpha$ collision is
 \begin{equation}
 l_{j,\mathrm{C}}=\frac{q_\alpha q_j}{4\pi\varepsilon_0\mu_j u_j^2}\label{l_j,C}
 \end{equation}
   with $u_j=|\mathbf{v}_\alpha-\mathbf{v}_j|$ the relative velocity between $j$ and $\alpha$-particle, and the  quantum impact parameter is $l_{j,\mathrm{Q}}=\lambda_q/4\pi  $ with  $\lambda_q=2\pi\hbar/\mu_j u_j$  the de Broglie wavelength~\cite{Tokamaks2}.
For the plasma status interested for inertial fusion, $e$-$\alpha$ collisions are in quantum domain and the corresponding Coulomb logarithm can be found from Ref.~\cite{Atzeni2004}:
\begin{equation}
\ln \Lambda_e=7.1-0.5\ln n+\ln T
\label{log_e}
\end{equation}
with $n$ in units of $10^{21}/\mathrm{cm}^3$ and $T$ in units of keV.
However, DT-$\alpha$ collisions is a little complicated, which can be treated either as classical or as quantum, depending on  the $\alpha$ particle energy and the DT ion temperature, while it is seldom discussed in previous publications.
Here, we take   the Maxwell distribution for DT ions and obtain the  average impact parameters for the DT-$\alpha$ collisions in  both classical and quantum cases. We first define a $T$-dependent critical energy $E_c$ to separate the classical domain from the quantum domain in calculating the Coulomb logarithm for the DT ions, which is fitted as:
\begin{equation}
E_c  =1663-\frac{6.9T}{1+0.001T}\text{ ,}
\label{E_c}
\end{equation}
where $E_c$ and $T$ are in units of keV.
Then, with the Maxwell-averaged $u_{\mathrm{DT}}^2=v_\alpha^2+\frac{3}{2}v_{\mathrm{th,DT}}^2$ and according to Eq. (\ref{log_j}, \ref{l_D}, \ref{l_j,C}), the Coulomb logarithm with quantum effect for the DT-$\alpha$ collision can be given as:
\begin{equation}
\ln \Lambda_{\mathrm{DT}}=\left\{
\begin{array}{c}
\ln \Lambda_{\mathrm{DT,C}}\text{, \ \ \ \ \ \ \ }%
E_\alpha < E_c \\
\ln \Lambda_{\mathrm{DT,Q}} \text{%
, \ \ \ \ \ \ \ }E_\alpha \geq E_c%
\end{array}%
\right.
\text{ ,}
\label{log_DT}
\end{equation}
where $\ln \Lambda_{\mathrm{DT,C}}$ and $\ln \Lambda_{\mathrm{DT,Q}}$ are:
\begin{eqnarray}
\ln \Lambda_{\mathrm{DT,C}}&=&7.25+\ln(E_\alpha+2.4T)+0.5\ln \frac{T}{n}  \label{log_DT_C}\text{ ,} \\
\ln \Lambda_{\mathrm{DT,Q}}&=&10.94+0.5\ln\frac{E_\alpha T}{n} \label{log_DT_Q} \text{ .}
\end{eqnarray}
Here, $E_\alpha$ and $T$ are in units  of keV, and $n$ is in units  of $10^{21}/\mathrm{cm}^3$. Nevertheless, the quantum effect on the Coulomb logarithm  for the DT-$\alpha$ collision is very small, within $3\%$, in the range interested for ICF study.

In the AM model~\cite{Atzeni2004,Tokamaks2}, the Coulomb logarithms for DT-$\alpha$ collisions is  written as
\begin{equation}
 \ln\Lambda_{\mathrm{DT,A}}=9.2-0.5\ln n+1.5\ln T\text{ ,}\label{log_DT_A}
 \end{equation}
which is obtained by taking $\frac{1}{2}\mu_{\mathrm{DT}}u_{\mathrm{DT}}^2=\frac{3}{2}T$ with an assumption that the relative velocity between these two kinds of particles is the same as the thermal velocity of the DT ions. However, this assumption needs to be modified at $v_\alpha\gg v_{\mathrm{th,DT}}$, especially for the newborn $\alpha$-particles.

\begin{figure}[htb]
\centering
\includegraphics[width=0.42\textwidth, clip]{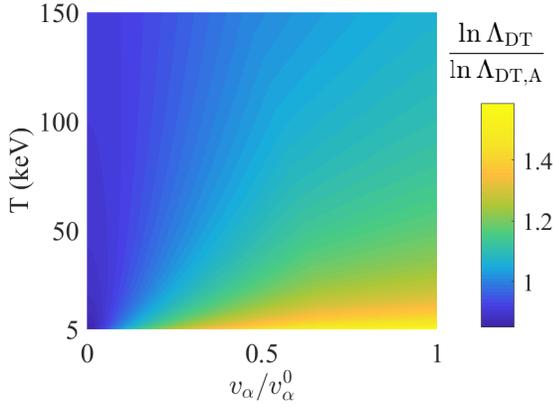}
\caption{(Color online) Color map of $\ln \Lambda_{\mathrm{DT}}$ to $\ln\Lambda_{\mathrm{DT,A}}$ ratio as a function of $E_\alpha$ and $T_{\mathrm{DT}}=T_e\equiv T$ from Eqs.~(\ref{log_DT}, \ref{log_DT_A}). }
\label{C_Q}
\end{figure}

In Fig.~\ref{C_Q}, we present the ratio of $\ln \Lambda_{\mathrm{DT}}$ to $\ln\Lambda_{\mathrm{DT,A}}$ in the map of $T$ and $v_\alpha$. As indicated, for $\alpha$-particles with $v_\alpha\gg v_{\mathrm{th,DT}}$  at 5 to 40 keV, $\ln \Lambda_{\mathrm{DT}}$ can be up to 1.6 times  larger than $\ln\Lambda_{\mathrm{DT,A}}$. As we know, a larger Coulomb logarithms can lead  to a stronger deposition of $\alpha$-particles and a weaker escape. Nevertheless,
with  all modifications considered in our model, the $\alpha$-particle  escape from our model is still  stronger than that from the AM model, as will be shown below.

As a summary,  we have considered the modifications of the $\alpha $-particle stopping by the DT ions and the electrons inside a  fuel plasma with the   Maxwellian average stopping weights, the relativity effect on electron  stopping weight
and the modified Coulomb logarithm of the DT-$\alpha $ particle collisions. Thus, the
variation of $\alpha $-particle energy within a displacement $ds$ inside a fully
ionized DT plasma can be rewritten from Eqs.  (\ref{dE_alpha}) and   (\ref{dE-Maxwellian}):
\begin{eqnarray}
\frac{dE_{\alpha }}{ds} &=& \left(\frac{dE_{\alpha }}{ds}\right)_{\mathrm{DT}}
+ \left(\frac{dE_{\alpha }}{ds}\right)_{e} \label{dE_alpha_approximation}
\end{eqnarray}
with
\begin{eqnarray}
\left(\frac{dE_{\alpha }}{ds}\right)_{\mathrm{DT}} = -\frac{m_{\alpha }}{m_{\mathrm{DT}}}\frac{e^{4}}{2\pi \varepsilon _{0}^{2}} \frac{n}{E_{\alpha}} \ln \Lambda _{\mathrm{DT}} g_{\mathrm{DT}},
\end{eqnarray}
and
\begin{eqnarray}
\left(\frac{dE_{\alpha }}{ds}\right)_{e} = -\frac{m_{\alpha }}{m_{\mathrm{DT}}}\frac{e^{4}}{2\pi \varepsilon
_{0}^{2}} \frac{n}{E_{\alpha}}   \xi \ln \Lambda _{e}g_{e*} .
\end{eqnarray}
Here,  we take $n_{\mathrm{DT}}=n_{e}\equiv n$ for a fully ionized ignited burning DT fuel, with $g_{\mathrm{DT}}$ and $g_{e*}$ calculated by using Eq.  (\ref{g}, \ref{g-e}),  relativity factor $\xi$ calculated by using  Eq.  (\ref{relativity}),
and the Coulomb logarithms $\ln \Lambda _{\mathrm{DT}}$ and $\ln \Lambda _{e}$  calculated by using Eq. (\ref{log_e}) and Eq. (\ref{log_DT_C}), respectively.
It shows a nonlinear deceleration
of $\alpha $-particles from Eq. (\ref{dE_alpha_approximation}), because it depends on   $E_{\alpha }$, $n$, $\ln \Lambda _{%
\mathrm{DT}}$, $\ln \Lambda _{e}$, $\xi$, $g_{\mathrm{DT}}$ and  $g_{e*}$, while these parameters  are strongly time and space dependent
inside a burning fuel plasma.

\begin{figure}[htb]
\centering
\includegraphics[width=0.42\textwidth, clip]{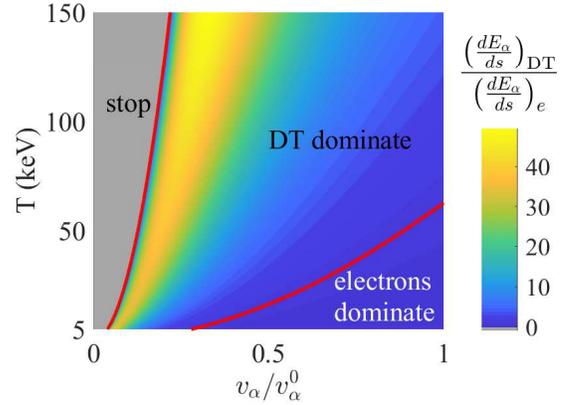}
\caption{(Color online) Color map of $\left(\frac{dE_\alpha}{ds}\right)_{\mathrm{DT}}$ to $\left(\frac{dE_\alpha}{ds}\right)_e$ ratio in the plane of $T$ and $v_{\alpha}/v_{\alpha0}$, with the  left red line  representing $\left(\frac{dE_\alpha}{ds}\right)_{\mathrm{DT}}=0$ when the $\alpha$-particle stops  and  the right red line representing $\left(\frac{dE_\alpha}{ds}\right)_{\mathrm{DT}}$ = $\left(\frac{dE_\alpha}{ds}\right)_e$ when the DT ions and the electrons contribute the same on stopping   the $\alpha$-particles. In the
middle part with $\left(\frac{dE_\alpha}{ds}\right)_{\mathrm{DT}}>\left(\frac{dE_\alpha}{ds}\right)_e$, the $\alpha$-particle stopping is dominated by the DT ions. In the right part with $\left(\frac{dE_\alpha}{ds}\right)_{\mathrm{DT}}<\left(\frac{dE_\alpha}{ds}\right)_e$, the $\alpha$-particle stopping is dominated by the electrons.}
\label{i_e_ratio}
\end{figure}

Here, it is interesting to compare the stopping contributions between the DT ions and the electrons.
Taking the same temperature for the DT ions and the electrons, $T_{\mathrm{DT}}=T_e\equiv T$, we obtain a map of  $\left(\frac{dE_\alpha}{ds}\right)_{\mathrm{DT}}$ to $\left(\frac{dE_\alpha}{ds}\right)_e$ ratio from Eq. (\ref{dE_alpha_approximation}), as presented in Fig.~\ref{i_e_ratio}. As indicated, it has two critical lines for the $\alpha $-particles. One is between the DT ion dominated deceleration phase and the electron dominated deceleration phase, defined as $\left(\frac{dE_\alpha}{ds}\right)_{\mathrm{DT}} = \left(\frac{dE_\alpha}{ds}\right)_e$. Another one is  between the  deceleration phase and the  stopping phase,  defined as $v_{\alpha} = v_{\mathrm{th, DT}}$. The deceleration   is mainly induced by the electrons   for  a newborn  $\alpha $-particle  at $T \lesssim 50$ keV. At a lower $T$, it has more slow electrons,
and the stopping from electron can dominate till to the case when the $\alpha $-particles are seriously  decreased.
At $T  >$ 50 keV, the deceleration of the $\alpha $-particles is fully dominated by the DT ions, because it has very few electrons with  velocity $v_e\leq v_\alpha$. For $T > $ 5 keV, as shown on the map, the deceleration phase of an  $\alpha $-particle  is dominated by the DT ions   when the $\alpha $-particle  is remarkably decelerated.
At a higher $T$, it has a higher $v_{\mathrm{th, DT}}$, thus an $\alpha $-particle stops deceleration and enters into its stopping phase with a higher velocity.

\section{Escape factor  of \texorpdfstring{$\alpha$}--particle }

\begin{figure}[htbp]
\centering
\includegraphics[width=0.3\textwidth, clip]{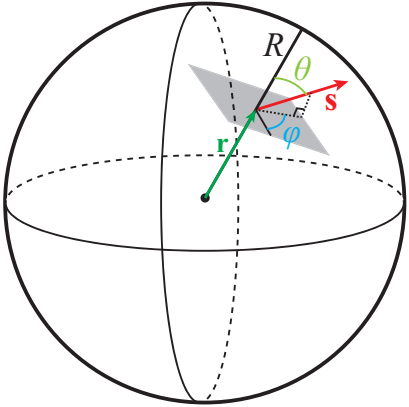}
\caption{(Color online) Diagram of   $\protect\alpha$-particles transferring
inside a spherical DT fuel of radius $R$. An $\protect\alpha$-particle  produced at $\mathbf{r}$ with a launching angle of $(\theta,\varphi)$    transfers along the red path $\mathbf{s}$.   Here, $\theta$ is the angle between $\mathbf{r}$ and $\mathbf{s}$, and $\varphi$ is an  azimuthal angle. The distance from $\mathbf{r}$ to the surface of the ball along direction $(\theta,\varphi)$ is $s_{0}(r,\theta ) = \sqrt{R^{2}-r^{2}\sin ^{2}\theta }-r\cos \theta $, which is independent of $\varphi$.}
\label{diagram}
\end{figure}

Escape of  the $\alpha $-particles   from a burning fuel is very important, which can seriously decrease the temperature of hot spot and thus fusion gain, and in addition, it can greatly influence a container design for future energy usage of inertial fusion.
From  Eq. (\ref{dE_alpha_approximation}),
by assuming that the launching of  $\alpha $-particles is isotropic and
the $\alpha$-particles do not change their moving directions, same as in previous works~\cite{Krokhin1973,Atzeni2004}, we can define an escape factor of the $\alpha $-particles escaping  from a spherical fuel plasma with radius of $R$:
\begin{widetext}
\begin{equation}
\eta_E=\frac{1}{4\pi V E_\alpha^0}\int_V d\mathbf{r}\int_{0}^{2\pi }d\varphi\int_{0}^{\pi } d\theta E_{\mathrm{esc}}(\mathbf{r},\theta,\varphi)
\sin \theta \text{ .}
\label{eta_t_eq26}
\end{equation}
\end{widetext}
Here, $V$ is the volume of the fuel ball, and the escaped energy for an $\alpha$-particle produced at $\mathbf{r}$ with a launching angle $(\theta,\varphi)$, as shown in Fig. \ref{diagram}, is
\begin{equation}
E_{\mathrm{esc}}(\mathbf{r},\theta,\varphi)=\left\{
\begin{array}{l}
0\text{, \ \ if\ \ }\exists \text{ }s\leq s_{0}\text{, }%
E_{\alpha }(s)\leq T_{\mathrm{DT}} \\
\\
E_{\alpha }(s_{0})\text{, \ \ if\ \ \ \ }E_{\alpha }(s_{0})>T_{\mathrm{DT}}%
\end{array}%
\right.
\text{ ,}
\label{E_esc}
\end{equation}
where $s_{0}$ is the
distance from point $\mathbf{r}$ to the fuel surface along direction $(\theta,\varphi)$, and the remained energy of this $\alpha$-particle after transferring within a length of $s$ is
\begin{equation}
E_\alpha(s)=\int_{\mathbf{s}}\frac{dE_\alpha(s')}{ds'}ds' \text{ .}
\label{E_s}
\end{equation}
Here, the integration path  is
\begin{equation}
\mathbf{s}=\mathbf{r}+s'\mathbf{k}(\theta,\varphi)\text{ with } 0\leq s'\leq s\text{ ,}
\end{equation}
where $\mathbf{k}(\theta,\varphi)$ is the unit vector in the direction $(\theta,\varphi)$.

The integration  Eq. (\ref{E_s})  is taken along the transferring trajectories $\mathbf{s}$ of all $\alpha$-particles by considering the variations of temperature  and density which are time and space dependent, as in Fig. \ref{diagram}.
As presented in Eq. (\ref{E_esc}), we consider that an $\alpha $-particle is fully stopped in the DT
plasma if the energy of this $\alpha $-particle reduces to $T_{\mathrm{DT}}$ at $%
s\leq s_0$, and in this case no energy of this $\alpha $-particle escaped. Or else, the remained energy $%
E_{\alpha }(s_0)$ of this $\alpha $-particle will be carried out and  lost from the DT fuel.

\begin{figure}[htbp]
\centering
\includegraphics[width=0.42\textwidth]{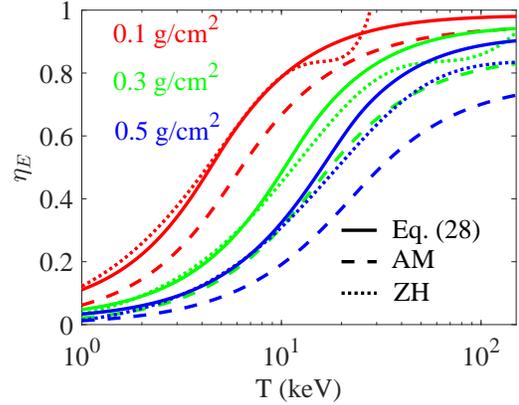}
\caption{(Color online) Comparisons among $\eta_E$ from   Eq. (\ref{eta_t_eq26}) (solid line),  the AM model (dashed line) and the ZH model (dotted line) for a burning fuel with $T$ varying from 1 to 100 keV at $\rho R = $ 0.1 g/cm$^2$ (red), 0.3 g/cm$^2$ (green) and 0.5 g/cm$^2$ (blue).}
\label{eta_t}
\end{figure}

In what follows, we calculate $\eta_E$ for a spherical DT fuel with uniform temperature and density, and compare $\eta_E$ obtained by using different models, for the parameters, radius $R=50$ $\mu$m, $\rho R$ = 0.1, 0.3, 0.5 g/cm$^2$ and $T$ = 1$\sim$100 keV. Presented in Fig.~\ref{eta_t} is comparisons among $\eta_E$ obtained from   Eq. (\ref{eta_t_eq26}), the AM model and the  ZH model. As shown, $\eta_E$  from the ZH model agrees well with that from Eq. (\ref{eta_t_eq26}) at $T < $ 10 keV, but it deviates from $\eta_E$ from Eq. (\ref{eta_t_eq26}) at  $T > $ 10 keV, for the reason that the expression of $\eta_E$ in the ZH model is fitted in the range of 1$\sim$10 keV~\cite{Zylstra2019}. Notice that  the AM model  gives the lowest $\eta_E$ as compared to the results from Eq. (\ref{eta_t_eq26}) and the ZH model, even at  $T\lesssim$ 5 keV.
The reason is that, the AM model calculates the escape factor with Eq. (\ref{Ratio_Ref}) from the KR model which ignores the deceleration of  the $\alpha$-particles induced by the DT ions, though it calculate the $\alpha$-particle range with Eq. (\ref{rho_l_Atzeni}) which takes   both DT ions and electrons into consideration. Thus, AM model gives a lower escape factor than    from   Eq. (\ref{eta_t_eq26}). This effect can be more remarkable at a higher temperature, when the DT ions dominate the stopping more significantly and the  relative effect is more seriously.

With all modifications of the $\alpha$-particle stopping by both DT ions and  electrons with their Maxwellian average stopping weights, the relativity effect on electron distribution and the modified Coulomb logarithm of DT-¦Á collisions, we   have a fitted expression of $\eta_E$ in Eq. (\ref{eta_t_eq26}):
\begin{widetext}
\begin{equation}
\eta_E=\frac{0.00593(\rho R)^{-1.174}T^{1.556}}{1+0.00385(\rho R)^{0.600}T^{1.316}+0.00547(\rho R)^{-1.180}T^{1.574}} \text{ ,}
\label{eta_t_fit}
\end{equation}
\end{widetext}
which can be applied to a burning fuel with  temperatures ranging from 1$\sim$150 keV and areal density ranging from 0.04$\sim$3 g/cm$^2$ with an accuracy within $\pm$ 0.02.

\section{\texorpdfstring Escape-effect on  hot-spot dynamics}

In this section, we study the $\alpha$-particle escape effect   on the hot-spot dynamics of an expanding burning plasma  by supposing  a DT fuel plasma which is an ideal gas composed with the DT ions and electrons with time-dependent uniform temperature  and  density for simplicity.
To focus on the escape effect, we  ignore the energy lost by radiation and by thermal conduction and
only consider  the self-heating  by the fusion products  $\alpha$-particles and the energy lost by expansion cooling via  mechanical work.   Thus, the dynamic
equations of the hot-spot can be written as
\begin{eqnarray}
\frac{dT}{dt} &=&\frac{2\pi }{9N}R^3E_{\alpha
}^{0}n^{2}\,\left\langle \sigma v\right\rangle(1-\eta_E)-\frac{2T}{R}\frac{dR}{dt} \text{ ,}  \label{T_dt} \\
\frac{dN}{dt} &=- &\frac{4\pi}{3}R^3n^{2}\left\langle \sigma
v\right\rangle \text{ ,}  \label{N_dt} \\
\frac{d^{2}R}{dt^{2}} &=&\frac{3NT}{m_{s}R} \text{ .}  \label{v_dt}
\end{eqnarray}
Here, $N$ represents the
total number of both DT ions and electrons. On the right-hand side of Eq.~(\ref%
{T_dt}), the first term represents the self-heating induced by $\alpha$%
-particles with escape effect considered, and the second term represent the expansion cooling of a burning plasma due to its mechanical work.
Eq.~(\ref{N_dt}) describes the number change of the DT ions and electrons
during the nuclear reaction, where we simply consider four particles are consumed in one reaction, including two DT ions burned and two electrons lost for
 electric neutrality. Eq.~(\ref{v_dt}) describes the expansion dynamics of the  burning
hot-spot   like that an ideal gas pushes a spherical piston outward with an acceleration of ${3NT}/{m_{s}R}$, where $m_s$ is the effective mass of the piston.
The reactivity of DT $\left\langle \sigma v\right\rangle$ is~%
\cite{Hively1983}
\begin{equation}
\left\langle \sigma v\right\rangle=9.1\times 10^{-22}\exp
\left(-0.572\left\vert \ln \frac{T}{64.2}\right\vert ^{2.13}\right) \text{ ,} \label{rate}
\end{equation}
where $\left\langle \sigma v\right\rangle$ is in unit  of m$^3/$s and $T$ is in unit  of keV.
The gain of the hot-spot can be defined as
\begin{equation}
G_H=\left(\frac{\Delta N}{4}E_{\mathrm{fus}}\right)/\left(\frac{3}{2}T%
_0N_0\right) \text{ .}  \label{G}
\end{equation}
Here, $T_0$ and $N_0$ are  temperature and particle number  of the hot spot at its stagnation phase, respectively, $\Delta N = N_0 -N$ is  the particle number difference of the hot spot  during its expansion phase. On the right-hand side of Eq.~(\ref{G}), the numerator is the total released
fusion energy with $E_{\mathrm{fus}}=17.6 $ MeV, and the denominator is the
thermal energy of the hot spot at stagnation phase which approximates to the mechanical work during the compression phase. For the indirect drive laser fusion, to calculate the fusion energy gain $G$, one should also consider the   absorbed laser efficiency $\eta_{\mathrm{aL}}$, the laser-to-X-ray conversion efficiency $\eta_{\mathrm{LX}}$, the hohlraum-to-capsule coupling efficiency  $\eta_{\mathrm{HC}}$ and the  rocket
efficiency $\eta_{\mathrm{rocket}}$. Then, the fusion energy gain $G$ can be written as:
\begin{equation}
G=\eta_{\mathrm{aL}}\eta_{\mathrm{LX}}\eta_{\mathrm{XC}}\eta_{\mathrm{rocket}%
}G_H \text{ .} \label{Glaser}
\end{equation}
Usually, it requires $G \ge 1$ for ignition.

\begin{figure*}[htbp]
\centering
\includegraphics[width=0.34\textwidth]{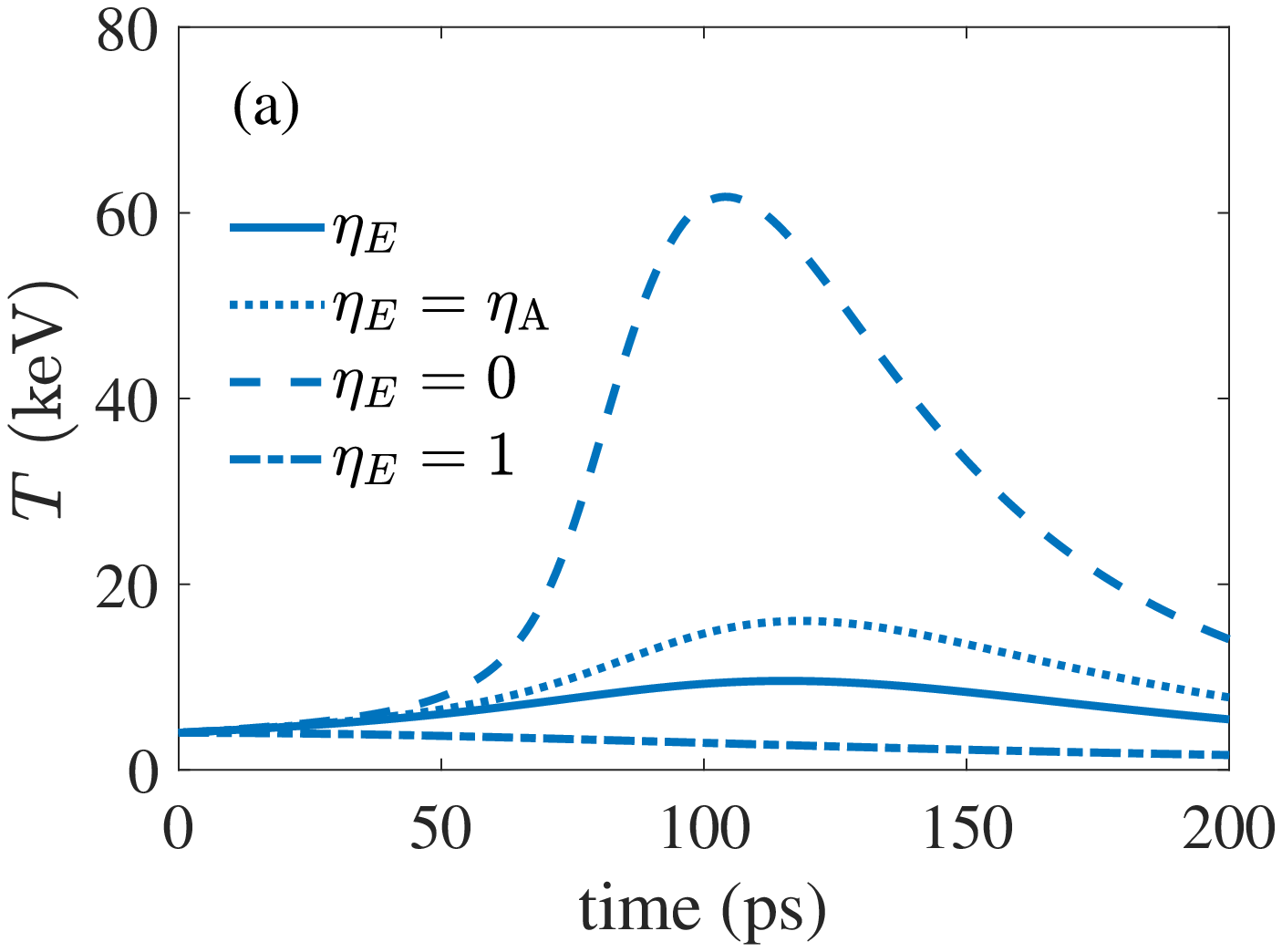} %
\includegraphics[width=0.34\textwidth]{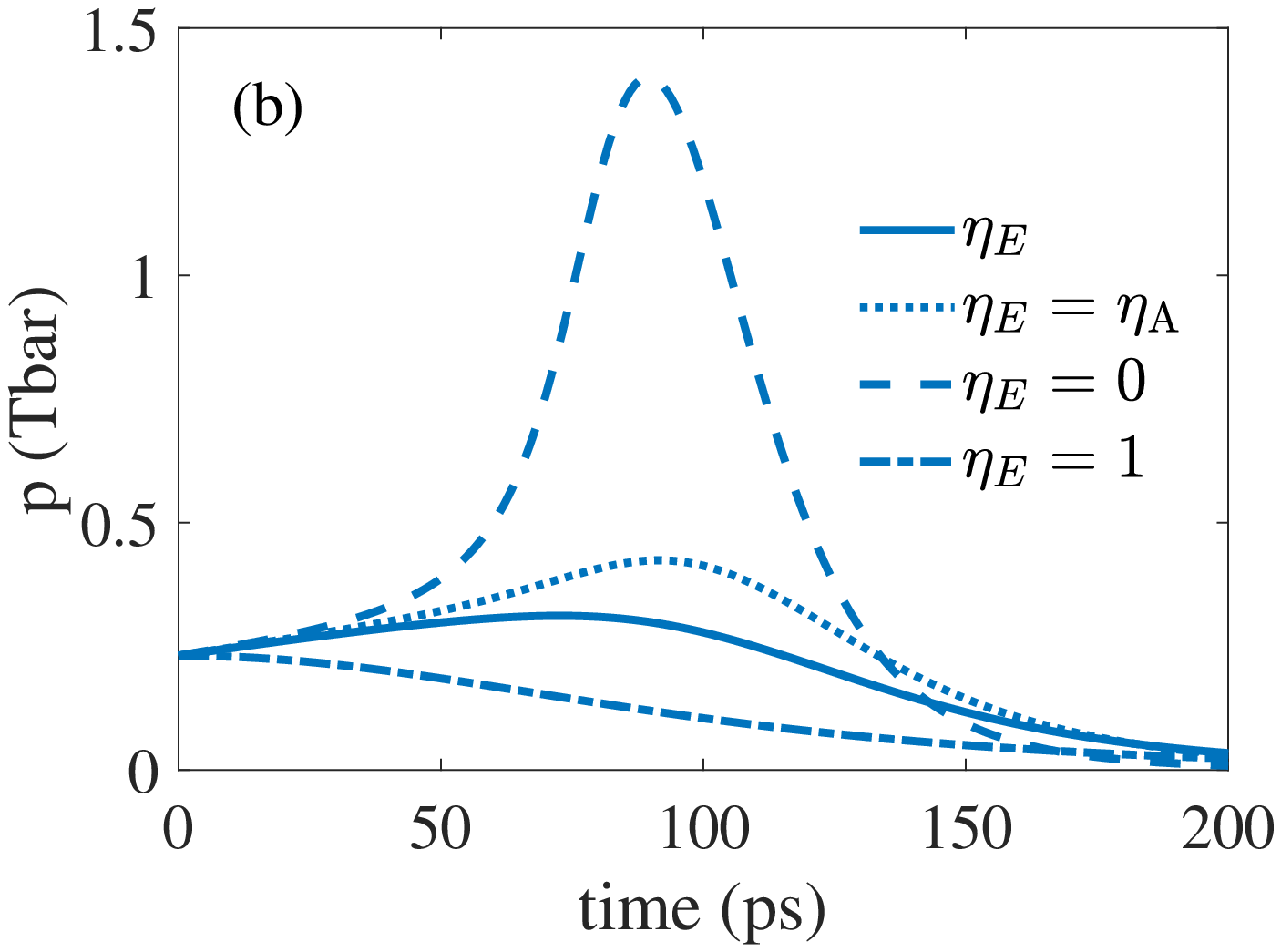} %
\includegraphics[width=0.34\textwidth]{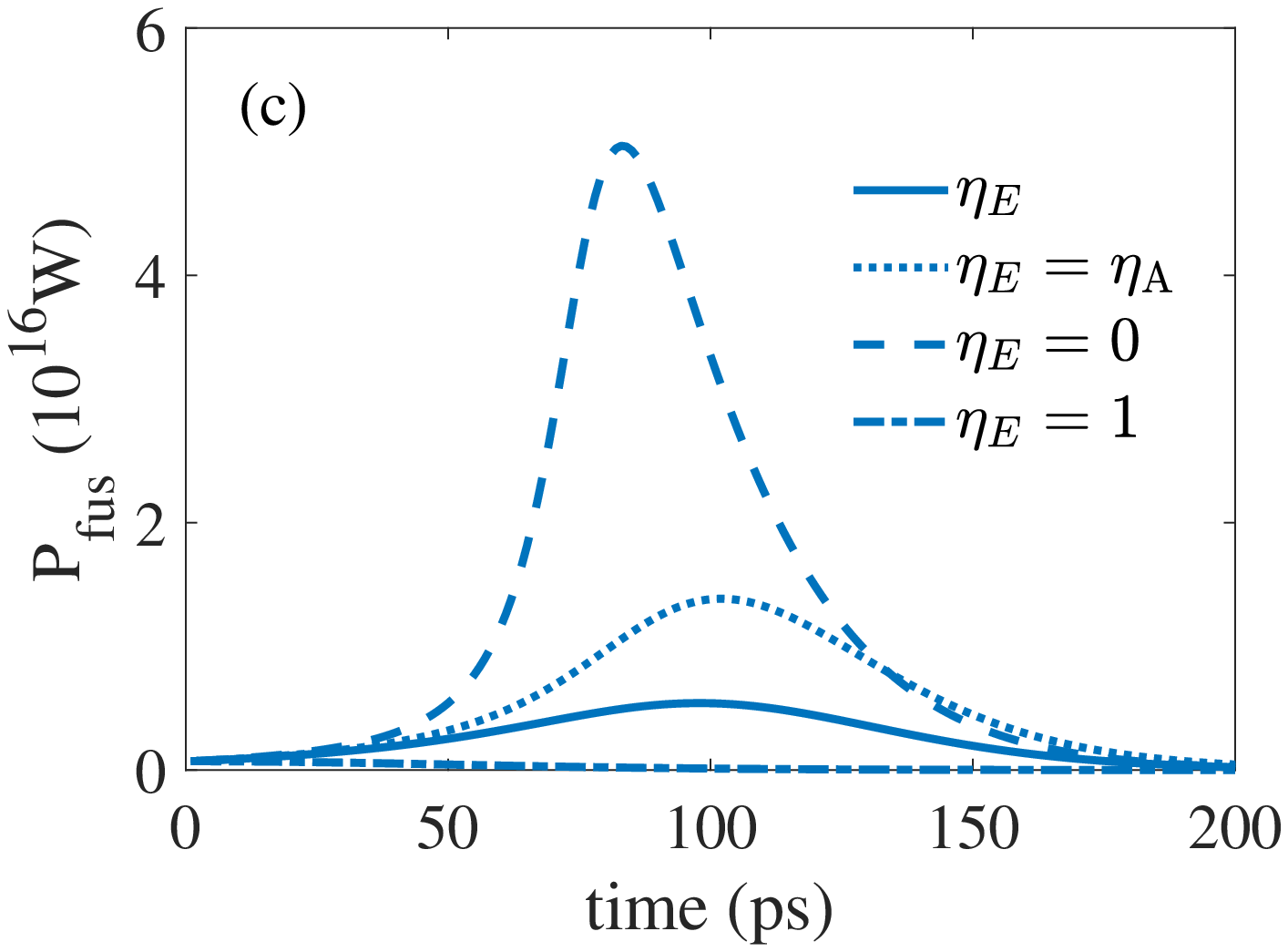} %
\includegraphics[width=0.34\textwidth]{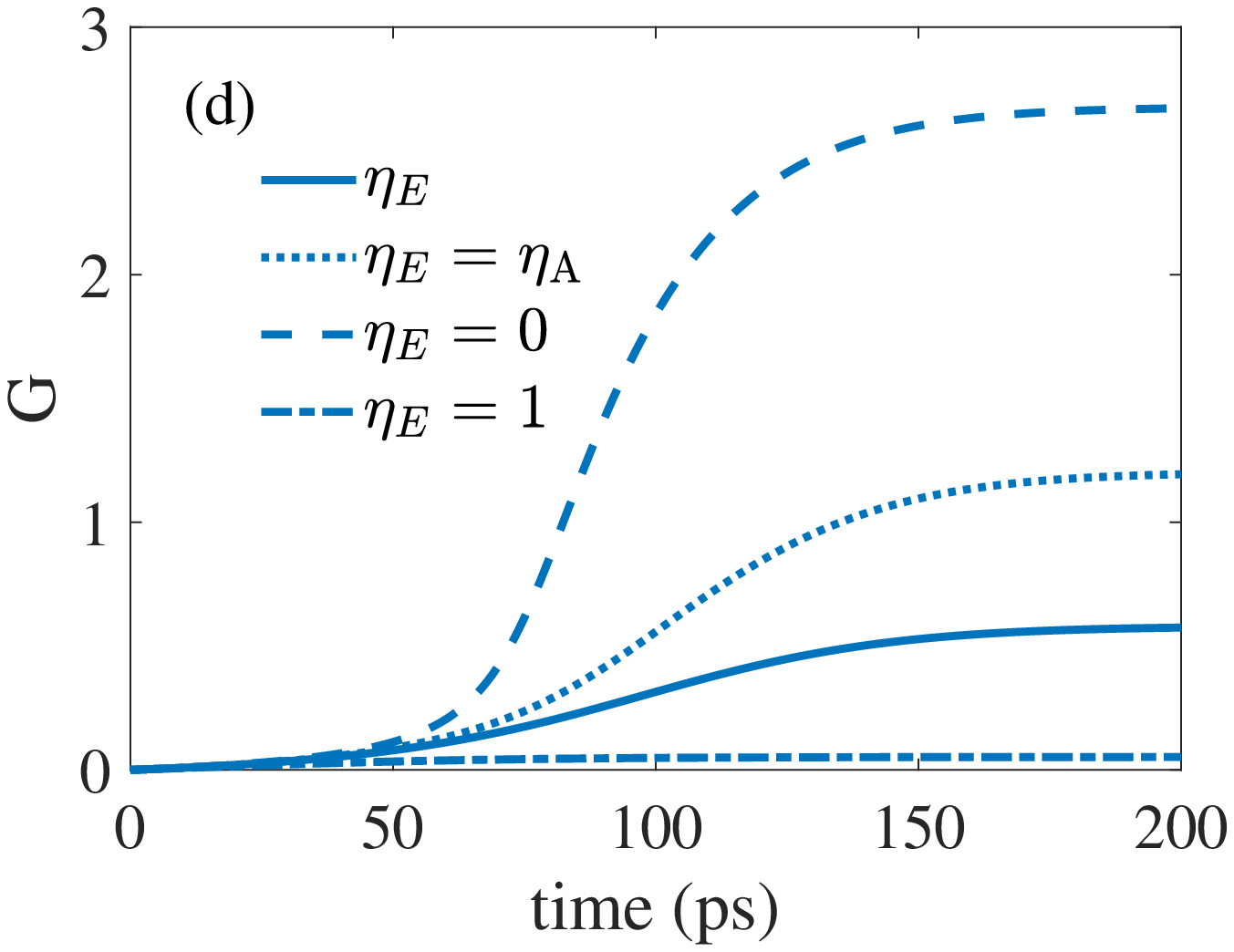}

\caption{(Color online) The temporal evolutions of (a) temperature $T$, (b) pressure $p$, (c) power for nuclear energy releasing $P_{\mathrm{fus}}$, and (d) energy gain $G$ of the hot-spot by using  $\eta_E$ from Eq. (\ref{eta_t_fit}) (solid line), $\eta_E = \eta_A$ [Eq. (\ref{Ratio_Ref})] from the AM model (dotted line), $\eta_E=0$ (dashed line) and  $\eta_E=1$ (dashed dotted line). }
\label{hotspot_1}
\end{figure*}

Considering a hot spot with initial radius $R_0=0.05$ mm, initial density $\rho_0=75$ g/cm$^3$, $T_0=4$ keV  and  an initial piston velocity   $dR/dt=0$ at the stagnation, we calculate the expanding  dynamics of this hot spot with above simple model. In this  calculation, the piston mass includes that of the cool fuel and the remaining ablator, which is  set as 10 times of the mass of hot-spot.
Presented in Fig.~\ref{hotspot_1} is temporal evolutions of the temperature, the pressure, the released fusion power and $G$ of the hot-spot   with  $\eta_E$ from Eq.~(\ref{eta_t_fit}).
For comparison, the hot-spot dynamics calculated with  $\eta_E = \eta_A$ from the AM model [Eq.~(\ref{Ratio_Ref})], $\eta_E$ = 1 (assuming all $\alpha$-particles escape out of fuel)  and $\eta_E$ = 0 (assuming all $\alpha$-particles deposit inside the fuel) are also presented.  Here, we take  $\eta_{\mathrm{aL}}\sim0.85$, $\eta_{\mathrm{LX}}\sim0.87$, $\eta_{\mathrm{XC}}\sim0.16$~\cite{Lan2016}, and $\eta_{\mathrm{rocket}}\sim0.17$~\cite{Lindl1995} in calculating $G$. Here, we don't consider the case with $\eta_E$ from the ZH model, because temperature range of the   escape factor expression from this model is limited up to 10 keV, much lower than what we concern for an ignited burning fuel.

As shown in Fig.~\ref{hotspot_1}, the  $\alpha$-particle escape-effect can  significantly influence  the hot-spot dynamics in burning phase and influence  the nuclear energy  released by fusion. First, the self-heating effect induced by $\alpha$-particles plays a key role in the inertial confinement fusion, which is a prerequisite for ignition. For the case with $\eta_E =1$, i.e., no $\alpha$-particle deposition, $G$ is about zero. Second, the self-heating is significantly  connected with the the escape of the $\alpha$-particles. At a smaller $\eta_E$, i.e., with less  $\alpha$-particles escaping and so a higher self-heating, the hot spot has a higher temperature, a higher pressure  and a higher fusion power. For example, the maximum temperature   reaches 63 keV and maximum pressure reaches 1.4 Tbar  at  $\eta_E = 0 $, while they are only  16 keV and 0.4 Tbar by using   $\eta_E$ from the AM model, and 9 keV and 0.3 Tbar by using $\eta_E$ from Eq.~(\ref{eta_t_fit}). Thus, it can seriously overestimate the self-heating by assuming all $\alpha$-particles are deposited. Third,  calculation of the escape factor of  $\alpha$-particles is  important in obtaining a more reliable hot spot dynamics.
In contrast to our model by using $\eta_E$  from Eq.~(\ref{eta_t_fit}), the case with $\eta_{\mathrm{A}}$  can  underestimate  the escape-effect, which leads to a higher temperature and a higher pressure  of the hot spot. Notice that $G$ is about 1.2 by using $\eta_E$ from the AM model, while it is only 0.6 by using $\eta_E$ from Eq.~(\ref{eta_t_fit}). That is to say,  it could ignite  by using the AM model, but the ignition  fails by using  $\eta_E$ which considers all modifications given in this work. In other words, the escape of  $\alpha$-particles can remarkably change the ignition condition.

Finally, it is interesting  to mention that the $\alpha$-particle escape effect can  also increase the gain   for a violent burning hot-spot with a temperature $\gtrsim$ 64 keV. The reason is that, the reactivity $\left\langle \sigma v\right\rangle$ of the DT fusion decreases with $T$ at $T\geq$ 64 keV~\cite{Atzeni2004} and meanwhile, the escape effect leads to a lower temperature and a lower pressure, and thus a slower expansion and higher density.  Because the  reaction rate is $n^2\left\langle \sigma v\right\rangle/4$, so the $\alpha$-particle escaping from an ignited  burning hot-spot with $\gtrsim$ 64 keV can lead to a colder fuel with a larger reactivity and a higher density and thus a higher gain.

\section{Summary}

We have studied the $\alpha $-particle escape from an   burning DT fuel with temperatures up to more than  tens keV by considering modifications of the $\alpha $-particle stopping by both DT ions and  electrons  with their   Maxwellian average stopping weights, the relativity effect on electron distribution, and the modified Coulomb logarithm of the DT-$\alpha $ collisions.
From our studies, it show that: (1) the deceleration is mainly induced by the electrons for a newborn $\alpha$-particle at T $\le$ 50 keV, while it is fully dominated by the DT ions for a seriously decelerated $\alpha$ particle or at $T > $50 keV; (2)  the relativity effect  can remarkably decreases the $\alpha$-particle stopping by electrons  at a high temperature, such as it can decrease by $28\%$ at 100 keV; (3) the modified Coulomb logarithm can be as much as 1.6 times of the ones in AM model.
We  gave a fitted expression Eq.~(\ref{eta_t_fit}) with consistent geometric treatment, to calculate $\eta_E$, which can be applied to the burning plasmas   of 1 to 150 keV and   0.04 to 3 g/cm$^2$ with an accuracy within $\pm$ 0.02. This expression can be used to estimate the $\alpha$-particle escape for a DT fuel with a uniform density and a uniform temperature  and with the same temperature  for the DT ions and the electrons. However, one should use the integration Eq. (\ref{eta_t_eq26}) to obtain a more reliable escape factor for the hot-spot which plasma status is  time and space dependent.
We  further discussed  the $\alpha $-particle escape-effect on the hot-spot dynamics  by comparing the calculation results with   $\eta_E$  from  different models.
As a result, the hot-spot dynamics of a burning fuel is strongly connected with the   escape of  $\alpha$-particles.
The escape factor of  $\alpha$-particles from our model is larger than   previously published results, which  can to a lower self-heating,  a lower temperature, a lower pressure, and thus leads to a lower energy gain. The escape of  $\alpha$-particles may fail the ignition for parameters near the ignition condition. However, for a violent burning with high temperature ($\gtrsim$ 64 keV), the $\alpha$-particle escape can increase the fusion energy gain.

\acknowledgments
This work is supported by the China Postdoctoral Science Foundation (under Grant No. 2019M650584).

\bibliographystyle{apsrev4-1}
%
\end{document}